\documentclass[10pt,letterpaper]{article}
\usepackage{authblk}
\usepackage{graphicx}
\usepackage{enumerate}
\usepackage{pslatex}
\usepackage{apacite}
\usepackage{array}
\usepackage{booktabs}
\usepackage{float} % Roger Levy added this and changed figure/table
\title{End-to-End Models for the Analysis of\\ \textit{System 1} and \textit{System 2} Interactions\\ based on Eye-Tracking Data}
 
\author[1]{Alessandro Rossi}
\author[1]{Sara Ermini}
\author[1]{Dario Bernabini}
\author[2]{Dario Zanca}
\author[1]{Marino Todisco}
\author[3]{Alessandro Genovese}
\author[1,3]{Antonio Rizzo}
\affil[1]{DISPOC, University of Siena}
\affil[2]{DSMCN, University of Siena}
\affil[3]{AIDILAB, Siena, Italy}

\date{}

\def\nwi{NWI}
\def\nwoi{NWoI}
\def\rwi{RWI}
\def\rwoi{RWoI}

\begin{document}

\maketitle

\begin{abstract}
\noindent While theories postulating a dual cognitive system take hold, quantitative confirmations are still needed to understand and identify interactions  between the two systems or conflict events. Eye movements are among the most direct  markers of the individual attentive load and may serve as an important proxy of information. In this work we propose a computational method, within a modified visual version of the well-known Stroop test, for the identification of different tasks and potential conflicts events between the two systems through the collection and processing of data related to eye movements. A statistical analysis shows that the selected variables can characterize the variation of attentive load within different scenarios. Moreover, we show that Machine Learning techniques allow to distinguish between different tasks with a good classification accuracy and to investigate more in depth the gaze dynamics.
\end{abstract}

\section{Introduction}\label{introduction}

Viewing is a complex activity, involving cognitive aspects, conscious and unconscious. It manifests itself through motor behavior that acquires salient information in the form of light radiation. When observing static images, this attentive activity exhibits rapid eye movements called \textit{saccades}, occurring between the so-called \textit{fixations}. During fixations, the eye remains fixed and the information is sampled. It is well known that the cognitive load of individual tasks may influence eye movements statistics~\cite{mcmains2009visual,connor2004visual,mathot2018pupillometry}, and in particular some variables like \textit{average fixation duration}, \textit{saccade length} or \textit{saccade velocity}, among others. For this reason, it seems reasonable to define techniques based on eye-tracking data in order to recognize recurring patterns related to the visual attention and identify the task that the subject is performing~\cite{klingner2010measuring,zagermann2016measuring}. Indeed, it has already been observed that the variation of the attentive load within different tasks affects the eye movements~\cite{castelhano2009viewing,tanaka19}.

%which turn out to be useful in applications such as monitoring attentive state of drivers~\cite{palinko2010estimating} or understanding truth telling and deception~\cite{wang2010pinocchio}.

%Exploring the distinctions between subjects called to solve tasks with different cognitive loads, it has been hypothesized and formalized the presence of two systems~\cite{kahneman2011thinking} that emerge in different circumstances: \textit{System 1} operates when the subject is called to solve a task in a fast, automatic or unconscious way; \textit{System 2}, instead, operates by slow, logical or conscious processes. 

%le diverse attività e carichi cognitivi op

In this work we analyzed the vision behaviour of subjects involved in a Stroop test~\cite{stroop1935studies} while performing different visual tasks, naming and reading, in order to explore possible effects on the human attention. The exploratory patterns are expressed through variables related to eye fixations and saccadic movements, since they are both influenced by processing difficulty~\cite{pollatsek1986inferences}.
The execution of each task requires to the subject different attention loads: reading is performed as a fast and automatic process, while naming the color of a word written with an ink color mismatching its semantics is a slow conscious activity ~\cite{kahneman2011thinking}. 
The delay in naming colors of incompatible words has been described as a conflict between the System 1 and the System 2~\cite{kahneman2011thinking}. This phenomenon is well-known in experimental psychology and several methods have been developed to test and measure it~\cite{jensen1965scoring,dalrymple1966examination,bench1993investigations}. 
%Within the naming task~\cite{cattell1886time}, 
%This phenomenon is known in experimental psychology as Stroop Effect~\cite{stroop1935studies}, and 
To this aim, we set up a visual version of the Stroop test during which we recorded the eye movements of 64 subjects, following the experimental protocol defined in~\cite{megherbi2018emergence}. The experiment involve two different tasks, defined as Naming and Reading, and two conditions, defined as "With Interference" and "Without Interference". 

The goals are (1) to verify the presence of recurrent visual behavioural patterns for different tasks and conditions through a statistical data analysis and (2) to generate automatic models which are able to identify in which task or condition the subject currently is involved. %modello mhmm dell'anno scorso (vedere se metterlo)

%The data analysis leads to the identification of statistics on eye fixations and saccades, extracted at trial level, which are characteristics of the different experimental conditions. Classification performances using machine learning techniques underline that, under different conditions given by the Stroop test, gaze data is separable. This provides information about visual behaviour as well as an automatic method of wide applicability to detect tasks in which the interaction between System 1 and System 2 is contentious.

The paper is organized as follows. The section "Method" describes the experimental protocol set up for stimuli presentation and data collection. In the section "Experiments" we provide a detailed description of the data pre-processing, Machine Learning techniques and metrics for evaluation of the results. Finally, in the "Conclusions" we discuss results and suggest direction for future works.

\section{Method}\label{method}
We set up a visual Stroop test to record eye movements of 64 Italian subjects (32 females and 32 males, average age = 30,2 $\pm$ 11,72) following the experimental protocol defined in~\cite{megherbi2018emergence}. During the test, the participants had to perform two main tasks: Naming and Reading. These tasks were both divided into two conditions: one "With Interference" and one "Without Interference". The images representing the four stimuli were created by modifying and translating in Italian the ones originally proposed in~\cite{megherbi2018emergence}. Stimuli were presented as 1024 x 768 pixels images, divided in an equally spaced 4x4 grid to generate 16 identical cells, representing interest areas. A single word was placed in the center of each cell. The four generated stimuli were composed by: 
\begin{itemize} 
    \item Reading Without Interference (RWoI) - Participants had to read the words on screen. The words "ROSSO" ("red"), "GIALLO" ("yellow"), "VERDE" ("green") and "BLU" ("blue") were all colored black.
    \item Reading With Interferences (RWI) - Participants had to read the words on screen. The words "ROSSO", "GIALLO", "VERDE" and "BLU" were coloured red, yellow, green and blue, with a mismatching between the shade used and the meaning of the word (e.g. "ROSSO" was never coloured in red).
    \item Naming With Interference (NWoI) - Participants had to name the color of the words on screen. In this case, the Latin letters were replaced by pseudo-letters constructed to match the real letters’ physical properties (height, number of pixels, and contiguous pixels) by reconfiguring their original characteristics~\cite{megherbi2018emergence}. The pseudo-words were colored red, green, yellow and blue. 
    \item Naming With Interferences (NWI) - Participants had to name the color of the words on screen. The composition of the screen followed the same principles used for the construction of the RWI condition. 
\end{itemize}
The presentation order of the conditions were randomized balancing the set of possibilities among participants. Eye movements were recorded by an EyeLink Portable Duo\footnote{\url{https://www.sr-research.com/eyelink-portable-duo/}} set to 500 Hz sampling rate and in Head-Free mode. Images were presented on a 17 inches display (1920 x 1080 pixels), placed perpendicularly in front of the participant at a distance from the eyes ranging in 46-52 cm. The experiment was executed individually only once and subjects unable to correctly read words and instructions on the screen were discarded. 
The light in the room and external noise are controlled by the experimenter. Written instructions are presented to the participants, followed by an oral explanation. The experiment is preceded by an initial unrecorded trial and a standard 5-point calibration. Between the instructions and the stimulus screens, a white screen containing a circular trigger located at the top-left corner of the task image were presented. Each trial began when the participant fixated the trigger for at least 100 ms. The trial is completed when the participant press the “space” key. During the execution, the experimenter annotated on an Excel spreadsheet any relevant information regarding the experience of the subject and possible technical issues (e.g. if the first calibration failed).

\begin{figure*}
          \centering
          \includegraphics[width=.8\linewidth]{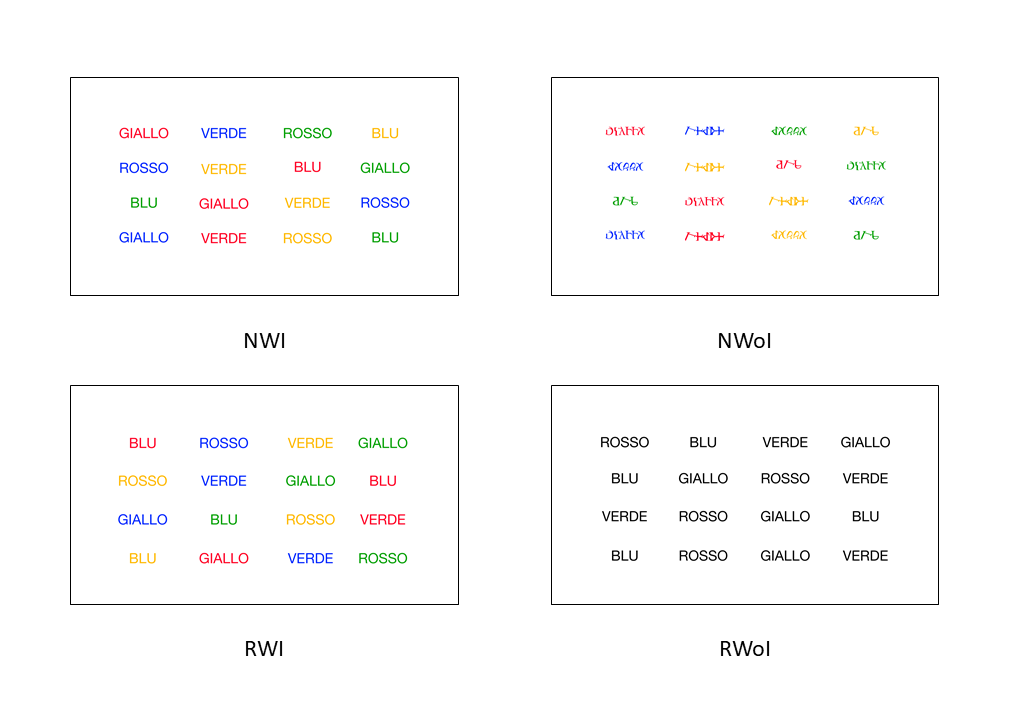}
                \caption{The four different stimuli presented to the subjects: Reading With Interference (RWI), Reading Without Interference (RWoI), Naming With Interferences (NWI) and Naming Without Interferences (NWoI). Each stimulus is presented in a different screen. Words are presented in Italian, because all participants are Italian mothertongue: "GIALLO" = yellow, "VERDE" = green, "ROSSO" = red and "BLU" = blue.}
                \label{fig:presentations}
    \end{figure*}
\section{Experiments}\label{results}

\subsection{Pre-processing}
Variables about eye movements were extracted by the software released with the eye-tracker device\footnote{\url{https://www.sr-research.com/data-viewer/}}. 
However, we found that a further cleaning process was necessary to improve the data quality.
Fixations that fall outside of the areas of interest were discarded, since they could be either due to an instrumental artifact or to a subject's activity which is not related to the task. 

Gaze data referred to head and tail of the experiment were also discarded. We refer to the head of a trial as the time until the trigger was activated. Indeed, the trigger dot was actually introduced to ensure similar initial conditions among subjects. In an analogous way, we refer to the tail of a trial as the time between the observation of the last word (the bottom right one) and the "space" key event which that concludes a trial. We empirically found that removing the last 5 fixations guarantees a good noise cleaning without filtering out any relevant fixations. 

A fixation threshold was used to discard fixations which are too short or too long, which could be due to noisy gaze points reads or to approximation errors introduced by the software. It is well known that meaningful fixations during reading tasks are within the range 100-400 ms~\cite{mcconkie2000eye,liversedge2000saccadic,majaranta2002twenty}. Because of the peculiarity of the task, we investigated a more specific method to select fixations of interest. Since fixations of interest for reading tasks are considered around 200-250ms, we dropped fixations below 200ms. For the longest ones, we applied an outlier detection methods, to select those samples with z-score~\cite{devlin1975robust} lower than 3. This approach appeared to be empirically valid, since allowed us to keep the maximum fixation duration in the interval of 800-1200 ms within subjects and experiments. %These values are higher than the ones considered in literature, but we would like to be more soft on this side since the fixation of maximum duration will be one of the variable considered in the analysis and a constant threshold would have cut the variability. 
This range is sensibly higher than the ones proposed in literature, but we avoided the use of a constant threshold in order to guarantee an additional degree of freedom so as to include the maximum duration of the fixations in a trial as one of the variables of interest. 

\subsection{Results}\label{res}

\begin{table}[t]
    \centering
    \resizebox{0.8\textwidth}{!}{
    \begin{tabular}{lccc}
       \multicolumn{1}{c}{} & \nwoi~vs. \rwoi & \nwi~vs. \nwoi & \multicolumn{1}{c}{\rwi~vs. \rwoi} \\
        \toprule
         Number of fixations &  0.285402	 & 0.000002 &	0.088730\\
         Average fixation length & 0.000052 & 0.027428 &	0.015745\\
         Maximum fixation length & 0.000606 &	0.035081 &	0.008078\\
         Horizontal regressions & 0.079062 &	0.000005 &	0.144568 \\
         Vertical regressions & 0.632652 &	0.000001 & 0.214819\\
         \cmidrule(lr){2-4}
        \multicolumn{1}{c}{} & \multicolumn{3}{c}{$p$-values}
    \end{tabular}}
    \caption{Significance $p$-values generated by the one-way ANOVA on fixations variables when comparing three pairs of tasks: \nwoi~vs. \rwoi, \nwi~vs. \nwoi~and \rwi~vs. \rwoi.}
    \label{tab:anova}
\end{table}

For each subject and stimulus (\nwi, \nwoi, \rwi, \rwoi), a set of statistical features related to eye movements were extracted: 
\begin{itemize}
    \item Number of fixations: total number of fixations.
    \item Average fixation length: the average of the duration among all the fixations.
    \item Maximum fixation length: the maximum of the duration among all the fixations.
    \item Horizontal/Vertical regressions: the number of times that the eyes step backward in their horizontal/vertical path (assumed left to right and up to down respectively), excluding the changes of line in the horizontal counting.
    \item Up/Down/Left/Right Frequency: the counting of saccadic movements in each direction, normalized by the total number of saccades.
    \item Minimum/Average/Maximum saccade duration: statistics about the duration of each saccade.
    \item Minimum/Average/Maximum saccade velocity: statistics about the estimated velocity of each saccade.
    \item Minimum/Average/Maximum saccade amplitude: statistics about the amplitude of each saccade (in degrees of visual angle).
    \item Minimum/Average/Maximum saccade distance: statistics about the distance of each saccade (in degrees of visual angle).
    \item Minimum/Average/Maximum saccade slope: statistics about the slope of each saccade with respect to the horizontal axis.
\end{itemize}

A first statistical hypothesis testing was performed by a standard one-way ANOVA \cite{fisher1992statistical}. The purpose was to assess the representativeness of the variables within different tasks and conditions. For each variable we analyzed the differences in the generated statistical distributions when comparing two different conditions, in order to verify the confidence that the considered measurement is sampled from two different populations from the compared groups. The test is repeated for three comparisons: \nwoi~vs. \rwoi, \nwi~vs. \nwoi~and \rwi~vs. \rwoi. Since the difference between Naming and Reading tasks is proven in literature~\cite{kahneman2011thinking}, the first test is a sort of control for the whole experimentation setting. However, we believe that exploring connections with visual attention and eye movements could be of interest too. The second and third comparisons are the focus of the presented work in order to asses if different attention levels, required to perform two different tasks, can be caught by variables related to eye movements.
In Table~\ref{tab:anova} we report the significance values $p$ obtained for in each comparison involving the variable about fixations. As we can see, the differences between Naming and Reading tasks are well represented by statistics about the duration of fixations (both average and maximum). In the second test, the effects of interference in Naming is highly expressed by the number of fixations and the eye regressions in both axis. This results appeared to be in agreement with the literature, since saccades regressions\footnote{Even if we are analyze fixations, their positions and, hence, possible regressions are directly related to saccadic movements.} are found to be more frequent and larger when the reader encounter some difficulties~\cite{pollatsek1986inferences,murray1988spatial}. The results in this first two tests was also confirmed by subjects general interviews, in which they admitted to perceive the Naming task as highly counterintuitive, especially in presence of interference. On the other hand, they confirmed to perceive the Reading task as more trivial, with little additional difficulty introduced by interference. This perception is also in agreement with our results, since the $p$-values for the \rwi~vs. \rwoi  are in general higher with respect to the other tests. Nevertheless, we observed that variables related to saccadic movements produce a lower level of significance when considered alone. Indeed, the minimum significance value $p=0.005$ was achieved by Average Saccades Duration when comparing \nwi~vs. \nwoi, but in few other cases we obtained a $p<0.01$.

Despite of this results, none of the variables produce, by itself, a satisfactory tasks characterization. Moreover, these kind of statistics does not directly provide a predictive model with good generalization performances when we need to infer new knowledge on unseen data. Indeed, a threshold model achieves poor performances, probably because of the high inter-subjects variability. Our claim is that more complex behaviors involving the dynamics of the attentive process or task specific gaze strategies can be captured by more complex (non-linear) models. Such models can take into account non-linear interactions among variables, possibly represented by means of hidden representations, partially overcoming the high variability. In the context of hypothesis testing, we applied Machine Learning techniques to assess statistical significance through a dual approach in which we evaluated the performances of selected learning models in classification tasks between two populations that we assume to be distinct~\cite{vu2018shared,oquendo2012machine,mjolsness2001machine}. We applied different machine learning techniques and evaluated the performances achieved on the collected dataset. Since the four stimuli are presented to each subject, our dataset consists of 64 examples per class, which can be too small to capture complex dynamics. However, to partially address this inter-subject variability, we exploited a specific normalization techniques. For each subject, we computed the mean of each variable within the four tasks, and subtracted it to the original values. These should mitigate individual effects on each tasks, and improve the final representativeness of the variables.

\begin{figure*}[t]
          \centering
          \includegraphics[width=1.2\linewidth]{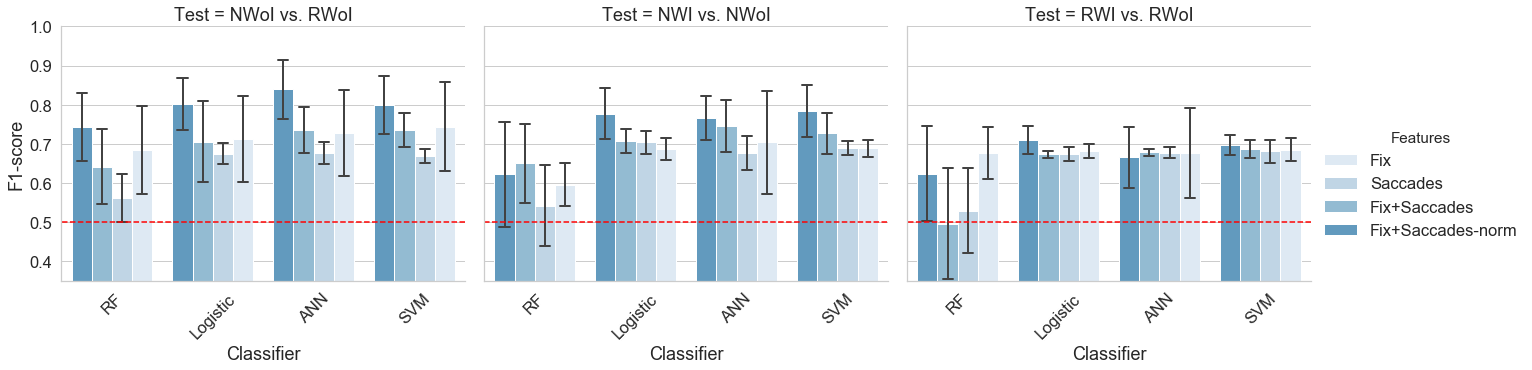}
                \caption{Classification performances in terms of F1-score. Each plot represents a different binary classification task, indicated in the title. Bars indicate the  average F1-score on a 5-fold cross validation setting, while the confidence interval  represented by black lines on the top of each bar indicates the standard deviation. Each group of contiguous bars refers to the performance achieved by the same classifier: SVM, RF, ANN and Logistic. Bar's color indicates the set of features fed as input to the classifier: \emph{Fix} (features related to fixations), \emph{Saccades} (features related to saccades), \emph{Fix+Saccades} (features related to fixations and saccades),  \emph{Fix+Saccades-norm} (features related to fixations and saccades with subject-wise normalization). The dashed red line indicates the score of the random baseline.}
                \label{fig:classification}
    \end{figure*}
    
We repeated the same tests investigated within the statistical analysis by setting up three separated binary classification test: \nwoi~vs. \rwoi, \nwi~vs. \nwoi~and \rwi~vs. \rwoi. We avoid a global 4-class test since the dynamic of the tasks are too complex to be modeled by a such small numbers of samples. We exploited the Scikit-learn~\cite{pedregosa2011scikit} Python software package to test four different classifiers~\cite{bishop2006pattern}:
\begin{enumerate}[i.]
    \item Logistic Regression (Logistic) is a statistical model that in its basic form uses a logistic function, applied to a weighted average of the input features, to model a binary dependent variable (the model prediction);
    \item Support Vector Machines (SVM) are supervised learning models used for binary classification. SVMs can learn non-linear separation surfaces by means of the so-called kernel trick, implicitly mapping their inputs into high-dimensional features space;
    \item Random Forests (RF) are an ensemble of decision trees based on bootstrapping. Different models are trained on a subset of samples and the final decision is taken by majority voting.
    \item Artificial Neural Networks (ANN) are a well known class of learning algorithms inspired by the biological neural networks; they are  based on a collection of units or nodes, called artificial neurons, connected by edges which represent the flow of information; edges are in fact numbers and represent the parameters of the model, typically learned by the back-propagation of an error signal with respect to the target.
\end{enumerate}
Our goal is to demonstrate that interactions between the two cognitive systems, System 1 and System 2, affect the gaze dynamics and can be detected by machine learning algorithms based on eye-related features. Eventually, we would like to find out that classification performances are consistently better than a random baseline in order to prove that the populations are intrinsically distinct. We performed a 5-fold Cross-Validation for each classifiers and computed the average of achieved F1-scores. This should guarantee that results do not depend on the choice of the test set, even if the relative high variability presented depends on the small size of the test (one single sample which is not classified correctly heavily affect the results). 

To improve the performances of Machine Learning algorithms, features were normalized in $[0, 1]$ (we found this method to slightly outperform z-normalization in our case). In addition, to investigate more in depth the information expressed by the features, we generated four sets of variables that we tested independently. 
\begin{itemize}
    \item \textit{Fix.} It was composed by the variables extracted from fixations, when filtering out fixations shorter than 200 ms. 
    \item \textit{Saccades.} It was composed by the variables extracted from saccades, aimed at capturing gaze dynamics and visual exploration schemes. 
    \item \textit{Fix+Saccades.} It is composed both from fixations and saccades features.
    \item \textit{Fix+Saccades-norm.} Since variables related to eye-movements are characterized by an strong inter-subject variability, for each subject we computed and subtracted the mean of each variable to generate the third set of features. These process was aimed at centering the variables distributions related to each subject around zero, translating the comparison among different subjects in a more comparable scale.
\end{itemize}
    
F1-scores achieved within the tests are reported in Fig.~\ref{fig:classification}. All the classifiers and features pairs are significantly above the random baseline, and in best cases above 0.8. These results corroborated the statistical analysis and the hypothesis that attention level influences gaze dynamics and those differences. Furthermore, these connections can be captured by an automatic classifier even at a small scale (i.e. with few training samples). Interestingly, a global trend is observed while considering different sets of input features. Combining features about fixations and saccades brings an improvement on the performances of each classifier, compared with the case in which separated features are exploited. This result also connects the attentive load to different exploration strategies of the visual scene. As already said, information about \emph{backward} saccades, are connected to more complex types of reasoning, typical of System 2, which are leaded by a need re-analysis or re-sampling of already visited portions of the scene~\cite{pollatsek1986inferences,murray1988spatial}. Moreover, a strong improvements have been achieved by applying a subject-wise normalization. This confirms that the analyzed scenario is highly affected by personal behaviors, but we showed that these effects can be mitigated by the application of standard statistical techniques. In general, we could observe that the Random Forests achieved performances which are considerably worse with respect to the others algorithms, sometimes even close to the random baseline. This could be due to the fact that decision tree are not capable to extract high-level correlations among variables, but most of all that the random sub-sampling negatively emphasize the inter-subject variability.
\section{Conclusions}\label{conclusion}
The experimental results supported the hypothesis (1) that different attentive loads present recurrent visual behaviors that can be characterized by a statistical analysis of variables related to eye fixations. Furthermore, these patterns can be modeled (2) by data-driven Machine Learning algorithms which are able to identify, with reasonable accuracy, the different conditions in which individuals are involved. We show that situations of conflict between System 1 and System 2 are captured by the gaze data and the statistical variables analyzed. The combining of features related to both fixations and saccades increases the accuracy of the classifiers. This suggests that subjects, among different tasks, use to implement task-specific schemes to regulate their gaze dynamics. We found that the exploited normalization techniques is useful when addressing wide inter-subject variability to improve the comparison among different individuals. However, these issues could be addressed more effectively by a large scale data collection to obtain more versatile Machine Learning models and more reliable results. 

Future research directions could include the integration in the analysis data related to pupillary response, since they are already proven to be connected to attentive and cognitive load~\cite{klingner2010measuring,mathot2018pupillometry}. This could help to explain more in depth connections among visual attention and eye movements, but also to develop more robust practical scenarios. Indeed, similar analysis turn out to be useful in applications such as monitoring attentive state of drivers~\cite{palinko2010estimating} or understanding truth telling and deception~\cite{wang2010pinocchio}.

\bibliographystyle{apacite}

\setlength{\bibleftmargin}{.125in}
\setlength{\bibindent}{-\bibleftmargin}

\bibliography{References}

\end{document}